\documentclass{article}
\pdfoutput=1

\usepackage[pdftex]{graphicx}
\usepackage{amssymb,amsfonts,amsmath}
\usepackage{authblk}
\usepackage{color}
\usepackage{fullpage}

\title{Frustration induced phases in migrating cell clusters}

\author[a]{Katherine Copenhagen}
\author[b,c]{Gema Malet-Engra}
\author[d]{Weimiao Yu}
\author[b,c]{Giorgio Scita}
\author[e]{Nir Gov}
\author[a,1]{Ajay Gopinathan}
 
\affil[a]{Department of Physics, University of California Merced, Merced CA 95343 USA}
\affil[b]{University of Milan, School of Medicine, Department of Oncology and Hemato-Oncology-DIPO}
\affil[c]{IFOM Foundation, Institute FIRC of Molecular Oncology, Milan 20139, Italy}
\affil[d]{Institute of Molecular and Cell Biology, National University of Singapore}
\affil[e]{Department of Chemical Physics, Weizmann Institute of Science, Rehovot, Israel}

\begin{document}
\maketitle

\begin{abstract}

Collective motion of cells is common in many physiological processes, including tissue development, repair, and tumor formation. Recent experiments have shown that certain malignant cancer cells form clusters in a chemoattractant gradient, which display three different phases of motion: translational, rotational, and random. Intriguingly, all three phases are observed simultaneously, with clusters spontaneously switching between these modes of motion. The origin of this behavior is not understood at present, especially the robust appearance of cluster rotations. Guided by experiments on the motion of two-dimensional clusters in-vitro, we developed an agent based model in which the cells form a cohesive cluster due to attractive and alignment interactions but with potentially different behaviors based on their local environment. We find that when cells at the cluster rim are more motile, all three phases of motion coexist, in excellent agreement with the observations. Using the model we can identify that the transitions between different phases are driven by a competition between an ordered rim and a disordered core accompanied by the creation and annihilation of topological defects in the velocity field. The model makes definite predictions regarding the dependence of the motility phase of the cluster on its size and external chemical gradient, which agree with our experimental data. Our results suggest that heterogeneous behavior of individuals, based on local environment, can lead to novel, experimentally observed phases of collective motion.

\end{abstract}

\section*{Introduction}
Collective motion is an emergent phenomenon in large groups of individuals where the motion can arise from purely local interactions. This phenomenon occurs across scales in systems ranging from bacteria to fish~\cite{Couzin2003,Zafeiris}.  Studies of such systems in the thermodynamic limit of infinite size have revealed a number of interesting features including long-range, scale-free correlations and a discontinuous phase transition~\cite{Guillaume2004}. These thermodynamic limit studies have spurred interest in hydrodynamic and mean field theories to describe such phenomena~\cite{Toner2005}. Finite groups display collective motion that closely model schools of fish or flock dynamics with boundaries. Their kinematics are characterized by unique behaviors including guidance by asymmetric boundaries~\cite{Wan2008} and the ability to simultaneously exhibit different phases of motion~\cite{Copenhagen2016}. Collective motion of groups was found to exhibit three distinct phases: running, rotating and random~\cite{Tunstrom2013a,Cheng2016}. In the running phase, the individuals are all more or less aligned, leading to a large translational velocity of the cluster center of mass. In the random or disordered phase, individual velocities are uncorrelated and there is very little overall motion of the cluster. In the rotating phase on the other hand, the cluster rotates as a whole around a common center. While the running and random phases have analogs in infinite systems, the mechanisms that can give rise to rotations are less clear. 
 
Through simulations, confinement has been shown, to be one mechanism that results in uniform populations of self propelled particles exhibiting rotational modes~\cite{Szabo2006,Leong2013,Camley2014a,Lober2015}.  Simulations of large groups of unconfined agents can also display rotational phases or milling states where the group rotates in a donut shape under certain conditions~\cite{Levine2000,Erdmann2005a,Couzin2002,Chuang2006,Carrillo2014}.  However to achieve these rotational milling states, the agents interact over a range up to tens of times the size of an individual agent, and form a low density `hole' at the center, where the defect in the velocity orientation field resides. Groups of cells have also been shown to display such rotations though it is unlikely that cells can interact much beyond their nearest neighbors~\cite{Mehes2014,Tanner2012}. This rotational motion has been studied both experimentally and using simulations for small groups of cells confined to different geometries~\cite{Brangwynne2000,Doxzen2013,Li2014a,Segerer2015}. Perhaps even more remarkably, unconfined cell clusters have also been observed experimentally to show transient rotational phases ~\cite{Malet-Engra2014}, and this behavior has been speculated to promote chemotaxis.

In this paper, we use an agent-based swarming model, which only allows short-range, nearest-neighbor interactions and unconfined space, similar to previous models found in the literature~\cite{Couzin2005,Belmonte2008,Copenhagen2016a}, to address the phenomenon of transient rotations in unconfined clusters. We show that a possible mechanism for driving cluster rotations is density-dependent cell propulsion.  This density-dependent propulsion may be caused by contact inhibition of locomotion (CIL), whereby cell protrusions are inhibited by the adhesions between cells~\cite{Zimmermann2016,Camley2016}. This causes cells at the cluster core, surrounded by other cells, to move slower than those at the edge of the clusters, which have a lower local cell density~\cite{Tarle2015}. This results in an outer rim of cells that move faster than central core ones and display stronger alignment interactions. We also find that decoupling the motion of the rim and central cells suppresses any rotational motion, suggesting that it is the coupling of two systems with different motilities (rim and core) that leads to rotational phases. Specifically, rotations arise in this model when the internal noise is such that the rim cells are in an ordered state with respect to velocity alignment, while the core of the cluster is disordered. The coupling of these two systems (rim around core) results in a frustrated state of the ordered rim being pinned by the disordered core. The whole coupled system is then able to relieve this frustration most effectively by existing in a rotational phase where the more ordered rim is able to move around the disordered core pinning it in place. This model successfully captures the dynamics of transitions between the modes of motion and proportions of time spent in each phase observed experimentally ~\cite{Malet-Engra2014}.  In the experiments, when the cell clusters are subjected, to a chemical gradient~\cite{Mittal2003,Rappel2016} there is an increase in the proportion of running phase, and a decrease in the rotational phase. This trend is also captured by our model when a chemical gradient is introduced. Furthermore, our model predicts an increase in the proportion of rotating phase with the size of the cluster, which we confirmed with experimental data. Taken together, our results suggest a novel form of frustrated interactions between behaviorally different parts of the same cluster that can lead to different collective dynamics -- a finding that may have applications beyond the context of cellular clusters.

\section*{Model}

Cell clusters are modeled as groups of particles that move with overdamped dynamics in two-dimensional continuous space (see SI-1). Cells are initially arranged in a circular disk, with velocities pointing in random directions. Cell velocities are determined by their internal self propulsion (with magnitude $p_i$), as well as physical interactions between cells, such as adhesions and collisions~\cite{Belmonte2008,Sepulveda2013}.  All of these interactions assume that cells communicate with each other by contact within a distance small enough to only include nearest neighbors. The cell diameter is selected from a Gaussian distribution, as uniform cell sizes lead to crystal lattice effects which are unlikely to exist in the experimental cell system (see SI-2). Finally, the velocities of the cells are subject to some uniform and uncorrelated noise ($\vec{\eta}$) due to random traction forces with the substrate and the random nature of the protrusions that cells use for propulsion. Cell positions are then updated according to their individually calculated velocities.

\begin{equation}
\vec{x}_i(t+\Delta t) = \vec{x}_i(t) + \vec{v}_i (t) *\Delta t
\label{pos}
\end{equation}

To determine the velocity of individual cells, a couple of interactions are taken into account. First, cells propel themselves in a direction ($\hat{n}$, eq.~\ref{prop}) determined by the memory of their own previous polarization and an alignment interaction with the mean orientation of neighboring cells, $\hat{V}$, with interaction strength $\alpha$.
 
\begin{equation}
\hat{n}=\frac{\hat{v}(t-\Delta t)+\alpha*\hat{V}}{|\hat{v}(t-\Delta t)+\alpha*\hat{V}|}, \: \hat{V}=\frac{\sum_{n.n.} \vec{v}_i}{|\sum_{n.n.} \vec{v}_i|}
\label{prop}
\end{equation}
 
\begin{figure*}[t]
\includegraphics[width=1\textwidth]{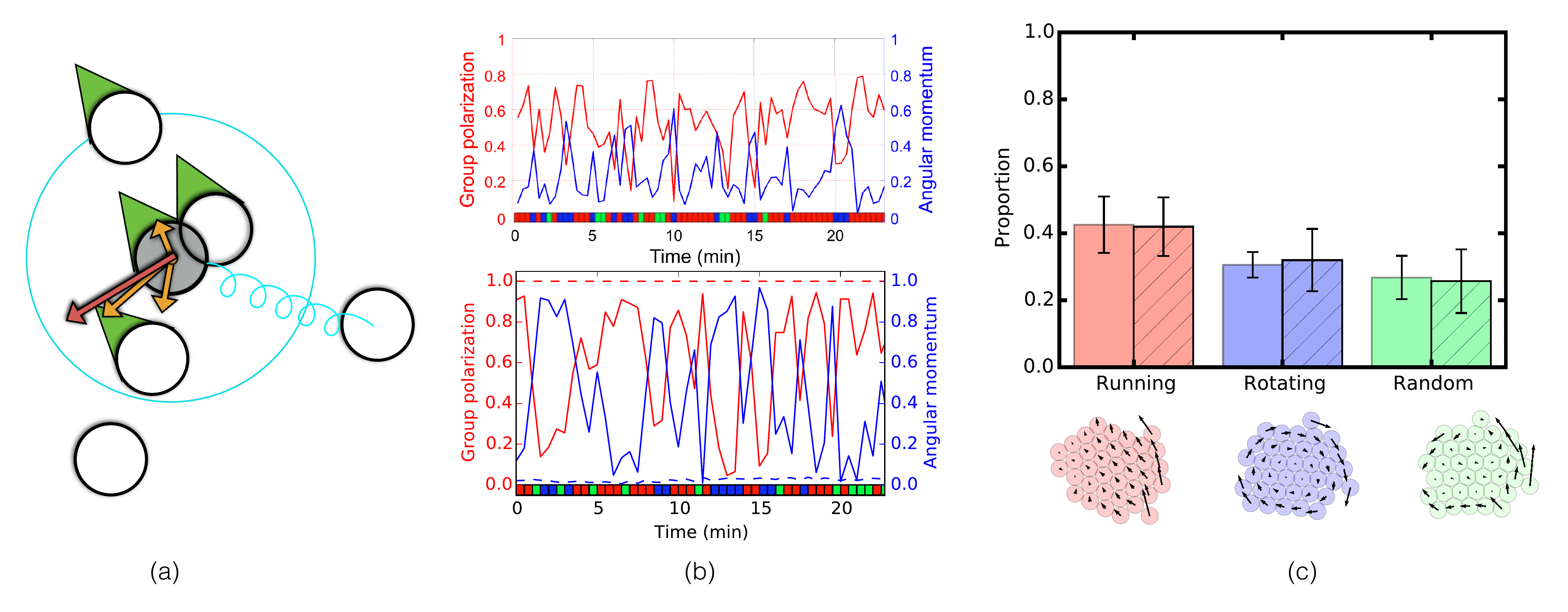}
\caption{(a) Schematic for the model. Green direction indicators show the direction of the neighbors of the gray cells, and the green indicator on the gray cell shows the alignment interaction ($\hat{V}$). The orange arrows show the Lennard-Jones interaction with each neighboring cell and the red arrow is the total LJ interaction ($\vec{LJ}$) on the gray cell. Finally, the blue spring denotes the cell-cell adhesion interaction ($\vec{S}$). Note that it only exists between the gray cell and its second nearest neighbors that do not have cells interrupting the path between them. (b) Time series of the group polarization and angular momentum of the cell cluster. The colors along the bottom axis show the phase of the system with time (running -- red, rotating -- blue, and random -- green) for experimental data (top) and simulations (bottom), for a uniform cluster (dashed) and a cluster with behavioral heterogeneity (solid, corresponding to the point marked in Fig.~\ref{prim8rs3nn}b). (c) The proportion of time that the cluster spends in each phase (simulations (plain) and experiments (cross-hatched)), along with a typical illustration of what each phase looks like in the simulations, with velocity vectors as black arrows. The cluster size for simulations is $N=37$ cells, while experiments are for an average cluster size of $50$.}
\label{schematic}
\end{figure*}

Cell velocities for each cell are calculated as arising from the forces described here and illustrated in Fig.~\ref{schematic}(a). The self propulsion magnitude is set by $p$.  Additionally cells experience volume exclusion and adhesion with neighboring cells, which are modeled as arising from a Lennard-Jones force ($\vec{LJ}$) and a spring-like interaction ($\vec{S}$). The spring force is longer range and acts past their nearest-neighbors, as long as it is not interrupted by other cells (Fig.~\ref{schematic}(a)). This adhesion force acts to maintain compact, cohesive, and roughly circular clusters.
 
 \begin{equation}
\vec{v}_i(t) = p * \hat{n}+\epsilon*\vec{LJ}+k*\vec{S}+ \vec{\eta}
\label{vel}
\end{equation}

\subsection*{Phase characterization}

We identify the mode of motion of the cell cluster by measuring the polarization ($\mathcal{O}$) and angular momentum ($\mathcal{A}$) 

\begin{equation}
\mathcal{O}=\frac{1}{N} \sum_{i=0}^N \hat{v}_i
\label{order}
\end{equation}
\begin{displaymath}
\mathcal{A}=\frac{1}{N}\sum_{i=0}^N \hat{v}_i \times \hat{r}_i
\end{displaymath}

We can also calculate the average expected velocity of the cells assuming perfect alignment, or perfect rotational motion (see SI-3).  We define the running phase to occur when the group polarization ($\mathcal{O}$) is greater than $0.5$, and the average cell velocity is over half the expected velocity assuming perfect alignment.  Similarly, the rotational phase occurs when the cluster angular momentum ($\mathcal{A}$) is greater than $0.5$, and the average cell velocity is over half the expected velocity assuming perfect rotation of the cluster.  Finally, the random phase is defined for all other combinations of $\mathcal{O}, \mathcal{A}$, and average velocity. These definitions are made to ensure consistency with the definitions in the experimental analysis~\cite{Malet-Engra2014}.  In the experiments, malignant B and T type lymphocytes were placed in a chemical gradient of CCL19, where they assemble into clusters and move towards higher CCL19 concentration. Automated analysis of video recordings of the cell clusters was utilized to extract velocity vectors of individual cells, which were then used to compute polarization and angular momentum as functions of time (see \cite{Malet-Engra2014} for details). Using the criteria described above, we can then label the phase of motion of the cluster for each time point. We then calculate the proportion of time that the cluster spends in each of the three phases throughout a simulation in a manner that allows for experimental comparison. 

\section*{Results}
\subsection*{Uniform cell clusters}
In the case where all cells within the cluster behave identically, the cluster remains in a single phase throughout the simulation. The dashed line in Fig.~\ref{schematic} (b, bottom) shows a time trace for a cluster in the running phase where the group polarization remains high and angular momentum remains low throughout. When compared to a time trace of the same quantities measured in experimental cell clusters (Fig.~\ref{schematic} (b, top), the features of the time traces are very different.  In the experiments, the group polarization and angular momentum fluctuate from high to low values corresponding to spontaneous transitions of the cluster between various phases of motion.  In the simulations, on the other hand, the cluster undergoes a transition from a running phase to a random phase only with increasing noise or decreasing propulsion.  Fig.~\ref{prim8rs3nn}a shows the proportion of time spent by clusters in each of the three phases plotted against noise and propulsion. The diagonal line of the transition between running at low noise and random at high noise is the well known noise-driven transition seen in Vicsek swarming models~\cite{Guillaume2004,Vicsek1995,Peruani2008}.

The running and random phases seen here are similar to those seen in experiments; however, the transition between the running and random phases in phase space is very sharp and there is very little overlap or mixing of the phases.  Experimentally, cell clusters are observed to spontaneously switch between running, rotating, and random phases which means that they coexist within this parameter space.  Alternatively one could say that cells change their internal parameters, such as the propulsion, $p$, or noise, $\eta$, so that they cross over the transition between the phases.  However, it is implausible for the entire cell cluster to change internal parameters in a coordinated way. Additionally, the uniform clusters show a very low level of cell rearrangement or fluidity within the cluster, whereas, in the experiments, cells are observed to move between the rim and the core of the cluster regularly. We therefore conclude that additional features of the real system must be incorporated into the model to recapitulate a rotational
phase and transitions between phases within a single set of parameter values, as well as large scale cluster rearrangement.

\subsection*{Introducing heterogeneous behavior}
An aspect of cellular behavior that is missing in this description is the possibility that cells may behave differently in different regions of the cluster, say the periphery or the interior.  Rim cells have increased propulsion compared to inner-cluster (core) cells due to reduced CIL, which causes cells adhered to other cells to form fewer protrusions than cells which have more open space around them. We implement this effect by scaling the propulsion with the number of neighbors, increasing for cells with fewer neighbors

\begin{equation}
p_i=p_{core}+\frac{3}{7}*(p_{core}-p_{rim})*(n_i-6)
\label{propmag}
\end{equation}

Here, $n_i$ is the number of neighbors around cell $i$.  $p_{rim}$ and $p_{core}$ are the propulsion of the rim cells (average of $3.67$ neighbors) and core cells (average of $6$ neighbors) respectively. A similar inverse relation between local density and propulsion (and therefore, alignment), was explored for a semi-infinite system in~\cite{Mishra2012}.

\begin{figure*}[t]
\includegraphics[width=1\textwidth]{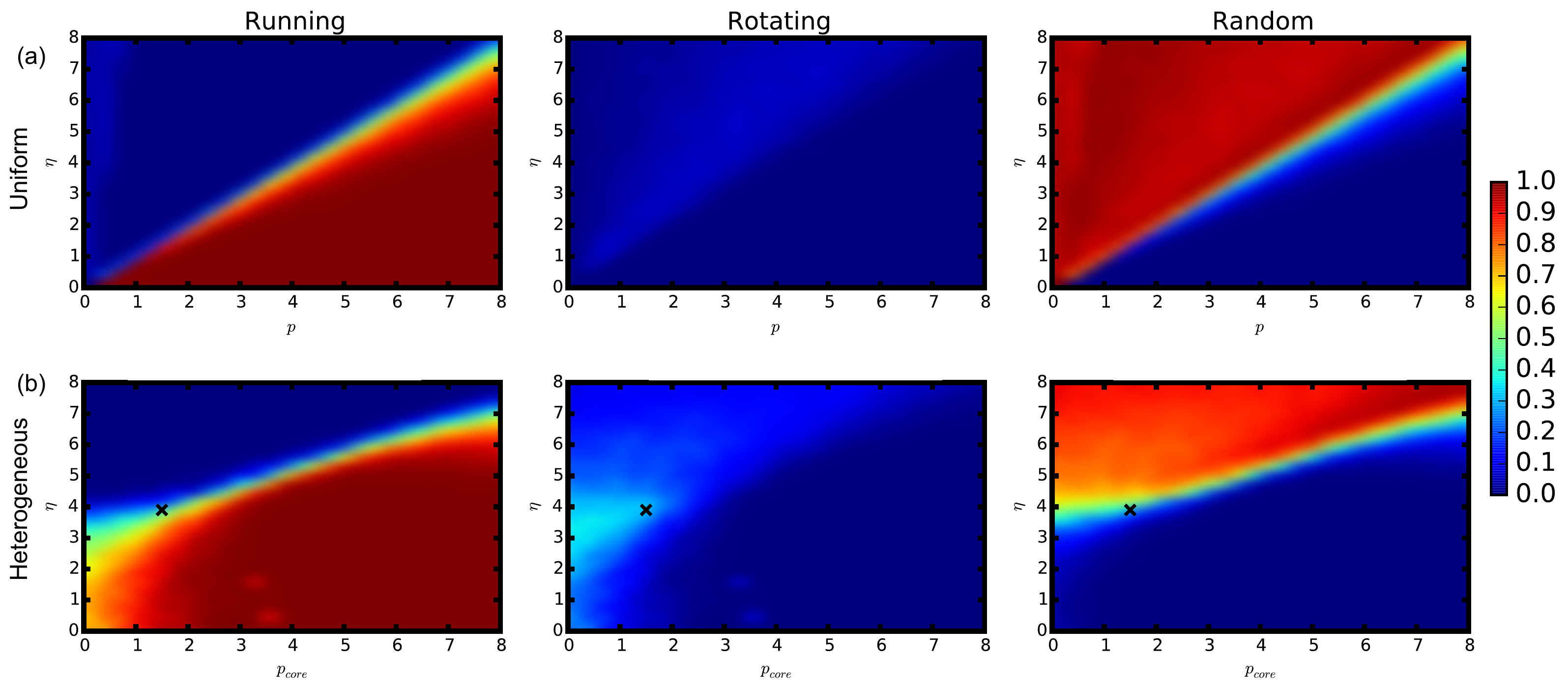}
\caption{(a) Proportion of time spent by the cluster ($N=37$ cells) in each of the three phases plotted against propusion $p$ and noise $|\vec{\eta}|$ for a cluster where all cells behave identically. (b) Phase diagram of the proportion of time spent in each of the three phases for a system with neighbor-number-dependent propulsion where the rim cells (those with $3.67$ neighbors) have a propulsion of $p_{rim} = 8$. $p_{core}$ is the propulsion of core cells (those with $6$ neighbors), and $\eta$ is the magnitude of the noise. The black `x' shows the point where the time series and phase proportions shown in Fig.~\ref{schematic} are taken.}
\label{prim8rs3nn}
\end{figure*}

This variation of cell propulsion causes the rotating phase to emerge, and to coexist with the other phases, as seen in experiments.  There is now a region in parameter space where there is a peak in the rotational phase at low values of $p_{core}$ and intermediate noise (Fig.~\ref{prim8rs3nn}b).  In this region, there are proportions of all three phases which are close to those seen in experiments.  The point highlighted in Fig.~\ref{prim8rs3nn}b is an example of a location in parameter space where the simulations closely match the experiments. The time series for the group polarization and angular momentum at this point in parameter space is shown in Fig.~\ref{schematic} (b, bottom), and agrees well with the experimental time series.  Furthermore, when the proportion of time spent in each of the three phases is compared to the experimentally measured values, they match very closely (Fig.~\ref{schematic}c). We next investigate, more closely, the mechanism that drives the rotating phase.

\subsection*{Cluster rim to core coupling}

\begin{figure*}[t]
\includegraphics[width=1\textwidth]{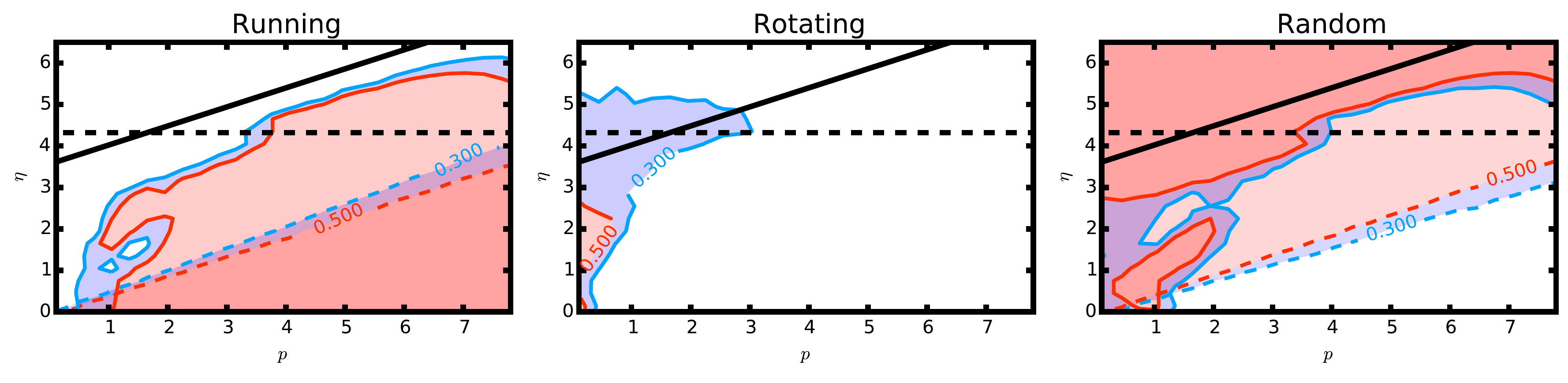}
\caption{Proportion of time spent by the system in each of the three phases as a function of propulsion $p$ and noise $|\vec{\eta}|$ for a ring of $18$ cells confined to a circle with propulsion $p_{rim}=p$. Dashed contour lines indicate regions (shaded) where the proportion of time spent in the corresponding phase exceeds 30\% (blue) and 50\% (red). Solid contour lines show the same contours but for the rim confined to a circle with $p_{rim} = 8$ coupled with a core of cells with $p_{core}=p$, and a full cluster size of $N=37$.  Note that the rotational phase only has non-zero values for the coupled system. The horizontal dashed line marks the noise value below which the rim alone would be ordered (greater than $30\%$ running phase) with $p_{rim}=8$, and the diagonal solid line marks the region above which a core with an average propulsion set by eq.~\ref{averagep} (with $p_{core}=p$) is disordered (greater than $30\%$ random phase).}
\label{rim}
\end{figure*}

The values of propulsion for individual cells in our simulation with heterogeneity are mostly close to $p_{core}$, except for those near the periphery where it rapidly climbs to an average of $p_{rim}$ (see SI-4). This prompts us to  consider whether the behavior of the system can be understood as arising from the coupling of two different systems -- a ring-like rim with a higher propulsion and a uniform core with lower propulsion. We first examine the rim cell system by confining a ring of cells to a circular shape and assigning them a fixed value of propulsion (= $p_{rim}$) independent of neighbor number. This results in the phase diagram shown in Fig.~\ref{rim}, where the contours are boundaries of regions where the proportion of time spent in the corresponding phase exceeds 30\% and 50\%. This phase space shares some characteristics with the uniform cluster phase diagram (Fig.~\ref{prim8rs3nn}a), such as the transition from a running to a random phase with increasing noise or decreasing propulsion, and the lack of a rotational phase. Due to the lower number of neighbors in the rim case, however, the slope of the diagonal running-random transition line is smaller compared to the case with a uniform cluster.

We next couple the rim cells to the core, resulting in a ring of cells confined to a circle with propulsion $p_{rim}= 8$, positioned around a core of cells with propulsion $p_{core}$. The black dashed line in Fig.~\ref{rim} shows the noise value below which a ring of cells with $p_{rim}=8$ would be ordered (or in the running phase more than $30\%$ of the time); while above the black solid line we expect a cluster with an average uniform $p$ given by eq.~\ref{averagep} to be disordered (random phase greater than $30\%$).  

\begin{equation}
p_{average} = \frac{p_{rim}*N_{rim}+p_{core}*N_{core}}{N}
\label{averagep}
\end{equation}

We notice that there is a triangular region between the dashed and solid black lines at low $p_{core}$ and intermediate noise where the rim should be in its ordered state while the core would be in the disordered state.  This suggests the possibility that when the rim is pinned in place by the random phase of the core cells it could relieve the frustration and maximize order by existing in a rotational phase moving around the core.

Indeed, when we couple the ordered rim to the random phase core in this parameter regime, we find a peak in the rotating phase (Fig.~\ref{rim}, solid contours).  The fact that the peak in the rotational phase does not exist for the rim or core alone but emerges when the two systems are coupled suggests that the rotational phase is driven by the coupling of the ordered rim to the disordered core.

There are some differences between the ring-disk confined system and the full unconfined model shown in Fig.~\ref{prim8rs3nn}b, but these are mainly due to affects arising from the confinement of the rim cells to a circle.  Indeed, relaxing the confinement of the rim cells, while still treating the cluster as two different coupled systems of the rim cells with higher propulsion around core cells with lower propulsion recovers the original, fully heterogeneous model phase space (see SI-5).  Taken together, these results suggest that it is the coupling between the disordered core and ordered rim that is the mechanism behind the cell cluster rotations seen experimentally.

\subsection*{Transitions into and out of rotating}
While we have shown that the running, rotating, and random phases can coexist within our model with heterogeneous neighbor-dependent propulsion, we have not yet examined the dynamics of the transitions between these phases. To do this, we quantify these transitions by monitoring changes in the overall topological properties of the phases, which are easier to track. In particular, note that in condensed matter systems, including active matter~\cite{Weber2014,Giomi2013}, phase transitions may be driven by the interactions and dynamics of topological defects~\cite{Duclos2014}. We first take a coupled rim-core cluster with rim cells confined to a circle and project the rim cell velocities onto the confining circle. We then identify a defect in this effectively one-dimensional velocity field as a point where the velocity projections switch directions. Note that the defects, as we have defined them, exist when there is no defect in the full velocity field of the entire cluster in its running phase and vanish when there in fact is a vortex in the cluster in its rotating phase (+/- 1 defect in the director field). 

\begin{figure*}[t]
\includegraphics[width=1\textwidth]{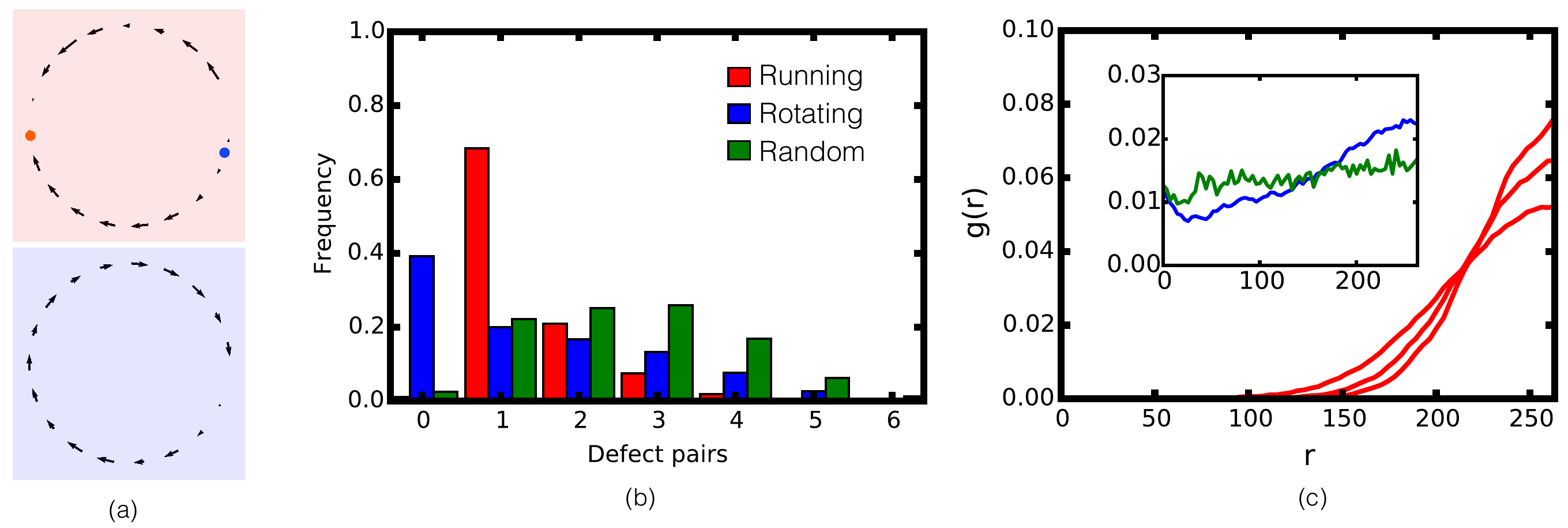}
\caption{(a) Velocities of the rim cells of a $37$ cell cluster which are confined to a circular shape projected onto the circle.  In the running phase (red), there are two defects of opposite signs in the velocity field, denoted by the orange and blue points. There are no defects in the rotating phase (blue). (b) The proportion of the number of defect pairs for each phase, with a peak at zero defect pairs for the rotating phase (blue), and one defect for the running phase (red). (c) The pair distribution function plotted against the separation between two defects when only one defect pair exists for parameters where the cluster primarily displays a running phase (note that g(r) is calculated over the whole simulation, independent of specific phases at any given point in time). (c - inset) The pair distribution function for points in parameter space dominated by rotating (blue) and random (green) phases.}
\label{defectsfluid}
\end{figure*}

Fig.~\ref{defectsfluid}a shows these defects for a cluster in the running phase and the rotating phase. In the running phase, the cluster has two defects of opposite signs at roughly opposite sides of the cluster. The formation and spreading apart of a defect pair coincides with the transition from rotating to running phase while the annihilation of the pair results in the running to rotating phase transition. By measuring the frequency of occurrence of any given number of defect pairs in each cluster phase, we investigate the correlation between the phase and number of defect pairs (Fig.~\ref{defectsfluid}b).  We see a large peak in the rotating phase for zero defect pairs and a peak in the running phase for one defect pair.  The random phase has a much broader peak around two or three defect pairs, suggesting that the random phase could occur when multiple defect pairs spontaneously form.

To observe the effective interactions of defects with each other, we calculate the pair distribution function ($g(r)$) for the spacing between two individual defects when a single pair exists. Fig.~\ref{defectsfluid}c shows the pair distribution function calculated over all time throughout a simulation, independent of what phase the cluster is in at any particular point in time. We see that for parameter values when the cluster is predominantly in the running phase, the pair distribution indicates that the two defects will repel and largely exist at maximum separation.  When the cluster is mostly in the rotating or random phase (Fig.~\ref{defectsfluid}c inset), there is a small peak at zero separation, implying that there may be a small effective attractive force between the defects if they get within one cell diameter of each other ($r_c\sim30$). At longer ranges, the interaction is repulsive, though the slope is much smaller for both of these cases than in the running case, so the defects only repel weakly, increasing their chance of annihilating and transitioning out of the running phase. Thus, system-wide parameters have a significant influence on the interactions between topological defects, which in turn controls the dynamics of the defects, the formation or annihilation of which are correlated with cluster phase transitions.

\subsection*{Cluster size dependence}

\begin{figure*}[t]
\includegraphics[width=1\textwidth]{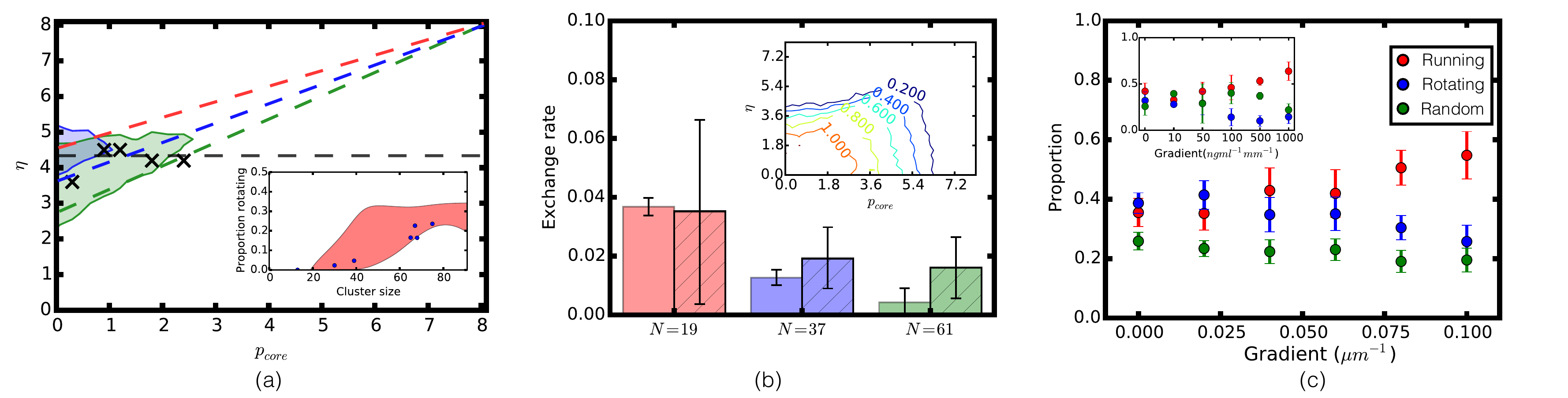}
\caption{(a) Proportion of time spent in the rotating phase by a system with neighbor-number-dependent propulsion as a function of $p_{core}$ and noise $|\vec{\eta}|$ for $p_{rim}=8$. The black horizontal dashed line marks the noise value below which the rim alone would be ordered (corresponding to the dashed transition line in Fig.~\ref{rim} with $p_{rim} = 8$).  The diagonal lines show the noise value above which a uniform system with average propulsion $p$ would be disordered (red: $N=19$, blue: $N=37$, and green: $N=61$).  The shaded regions are where the clusters spend at least 30\% time in the rotational phase, with the same color scheme (blue: $N=37$, and green: $N=61$; note that there is no red shaded region).  (a - inset)  The dependence of the proportion of rotating phase on system size. The shaded red region shows the range of dependence for a spread of parameter values marked with black crosses in the main figure.  The experimental measurements are shown as the blue points. (b) The fluidity of the cluster measured as the rate of exchange between the core and rim cells of the cluster, for several systems sizes, for both simulations (plain bars), and experimental data (cross-hatched bars). (b - inset) Contours for the fluidity of the cluster over the $p_{core}$-$\eta$ parameter space.  (c)  Simulated proportion of each phase (see legend) plotted with increasing chemical gradient ($g c' r$ in eq.~\ref{grad}, where $r$ is the cell diameter), along with experimental data in the inset. The concentration gradient of chemokine in the experiments is measured in (ng/ml)/mm and shown on the x-axis for the inset.}
\label{sizedep}
\end{figure*}

We next examine the effect of cluster size on the phase diagram. Fig~\ref{sizedep}a shows regions of the parameter space with a proportion of rotating phase greater than 30\%  for different cluster sizes.  Compared to the predictions of the simulations regarding the proportion of the rotating phase, the predictions regarding the running and random phases have a very large spread across the parameters tested. Thus, we focus on the more significant rotational phase shown here (see SI-6 for other phases).  Again, the dashed black line shows the noise value below which the rim alone would be in an ordered phase, while the colored dashed lines show the transition between the running and random phases for a uniform system of average $p$ (eq.~\ref{averagep}), for each different system size. Our results indicate that larger clusters have a higher proportion of rotational phase while smaller clusters are less likely to rotate. 

 This is consistent with the idea that the coupling-induced rotational phase only exists in the area of phase space where the rim propulsion would result in an ordered state (running phase greater than $30\%$ of the time; below the dashed black line) while the average $p$ of the whole cluster would lead to a disordered state (random phase greater than $30\%$ of the time; above the colored dashed lines for each size). Larger systems have a higher proportion of core cells and therefore will have a larger proportion of the phase space where the average $p$ results in a disordered cluster (because $p_{core}$ is always lower than $p_{rim}$) while the rim remains ordered, leading to a larger overlap of the two and a more stable rotational phase. Fig.~\ref{sizedep}(a inset) shows the comparison between experimental and simulation values, where the blue points are the experimentally measured proportions of rotating phase against cluster size, and the red shaded region shows the size dependence of the simulations with the parameter values marked by black crosses in Fig.~\ref{sizedep}a. Although these parameter values were chosen based solely on the proportions of all phases exhibited by a cluster of size $N=37$, the dependence on system size seen in the simulations is very similar to that of the experiments, further supporting the idea that the rim/core coupling is a likely mechanism for experimental cell cluster rotations. 

\subsection*{Cluster fluidity}

Exchanges between the periphery and the interior of the cell clusters were proposed to have an important functional role in exposing `fresh' cells with unsaturated receptors to the chemical gradient ~\cite{Malet-Engra2014}. To examine this feature of our clusters, we look at the fluidity of a cluster as measured by the rate of exchange between core and rim cells.  A rim cell is defined as any cell with an exposed edge larger than a single cell diameter. We then average the number of cells which switch between the rim and the core in each time step to get a measurement for the cluster fludity.

We measure this exchange rate for the original, fully heterogeneous neighbor-number-dependent propulsion model (phase space in Fig.~\ref{prim8rs3nn}b). Fig.~\ref{sizedep}(b inset) shows contours for the fluidity across parameter space. It turns out that the fluidity of the cluster is significantly higher when the cluster shows a majority of running or rotating phase compared to the random phase. This is presumably due to the fact that the large scale rearrangements or rim/core cell exchanges happen when the rim cells move past the slow moving core cells and then mix back into the cluster. This is why the fluidity drops for high $p_{core}$ values, approaching $p_{rim}$.   Fig.~\ref{sizedep}b shows the dependence of the rim/core exchange on cluster size for the simulations, as well as the experiments (cross-hatched bars).  The trend of decreasing rim/core exchange with increasing cluster size is more dramatic in the simulations, but is maintained between both experiments and simulation, further supporting a rim/core coupling as the mechanism for rotational phases in cell clusters.

It should be noted that the simulations slightly overestimate the proportion of the running phase in smaller clusters at the expense of the random phase. We speculate that this might be a signature of cellular clusters maintaining a roughly constant effective $p_{rim}$ across cluster sizes, perhaps by sensing curvature. In this case, our simulations are essentially overestimating the effective $p_{rim}$ value because the rim cells in smaller clusters have fewer neighbors, and $p$ increases linearly with decreasing number of neighbors. This would result in a slight overestimation of the running phase with decreasing cluster size at the expense of the random phase seen in the experiments (SI-7). Similarly, for large cluster sizes, the simulations underestimate the effective $p_{rim}$, leading to a higher proportion of the random phase at the expense of the running phase, as well as an underestimation of the rim/core exchange in simulations due to the fact that about $60\%$ of the exchanges that take place occur during the running phase in both experiments and simulations.

\subsection*{Cell clusters in a chemical gradient}

Cell clusters can chemotax robustly up a chemical gradient~\cite{Malet-Engra2014} and it has been shown that such a collective chemotactic motion can be obtained by cells at the cluster rim having a propulsive force normal to the surface of the cluster with a magnitude that depends on the local concentration of the chemokine~\cite{Malet-Engra2014,Levchenko2002,Camley2015b}.  We are interested in how such a gradient would affect the proportion of time the clusters spend in each of the different phases. To implement a chemical gradient into our model, we introduce an additional term into the calculation for the cell propulsion direction, $\hat{n}$, by replacing $\alpha*\hat{V}$ by $\alpha*\hat{V}+ \vec{g}$  in eq.~\ref{prop}), where

\begin{equation}
\vec{g}_i = g c' y \sum_j^{p. a. n.} \vec{f}_{j}
\label{grad}
\end{equation}

and the sum $j$ is over each distinct pair of adjacent neighbors of cell $i$.  $\vec{f}_j$ is a vector pointing in the direction bisecting the angle subtended by the centers of the cells of the neighbor pair at the center of cell $i$, with a magnitude equal to the arc length between the two neighbors (see SI-1.2). Here, $g$ reflects the strength of the influence on propulsion direction from the chemokine gradient per unit distance, of exposed cell edge arclength, $c'$ is the change in chemokine concentration per unit distance and $y$ is the distance (in microns) from the $0$ ng/ml concentration point. This results in a gradient force in the direction of the most vacant region around a cell, with a magnitude proportional to the size of the vacancy, causing a large outward force on rim cells and negligible force on core cells, resulting in an overall upward drift due to the unbalanced forces~\cite{Malet-Engra2014}.

We introduce such a chemical gradient to a system with running, rotating and random proportions close to those measured experimentally in the absence of a chemical gradient. We find that the gradient leads to cluster motion up the gradient as anticipated, and observed experimentally. We then measure the changes in the proportion of phases as a function of the gradient (shown in Fig.~\ref{sizedep}c). With increasing gradient, we find an increase in the proportion of the running phase and a decrease in the amount of the rotating phase, similar to what is seen experimentally, (Fig.~\ref{sizedep}c inset) while the random phase proportion stays essentially unchanged.  

\section*{Discussion}

Cell clusters exhibit running, rotating, and random phases in experiments. We have identified, using a theoretical model, the possible cause for the coexistence of these phases, in a cluster that has only short range cohesive and alignment interactions. It is based on the tendency of cells at the cluster rim to have increased propulsion due to less contact inhibition, and therefore stronger alignment interactions, compared to cells at the core of the cluster.  We have identified a likely mechanism by which this increased propulsion can lead to rotational phases in cell clusters.  

This effect involves the effective existence of two different systems within the cell cluster -- a high propulsion, ordered rim system and a low propulsion, disordered core. When these two systems are disconnected, there is no significant rotational phase present in either of them. However, when they are coupled together the rotational phase appears robustly, indicating that it is the coupling of these two systems that leads to the observed rotations. We find that the ordered rim is capable of dragging the disordered core with it, resulting in a solid-body-like rotation of the entire cluster (see SI-8). This behavior, whereby the ordered phase induces large-scale coherence in the adjacent disordered phase, is somewhat reminiscent of the coupling between super-conducting and normal metals (proximity effect)~\cite{Tinkham1996}.

Our simulations exhibit multiple features that are seen experimentally, including spontaneous transitions between the different phases, the cluster size dependence of the proportions of phases and the fluidity, as well as the response to chemical gradients. Our results indicate that larger clusters show an increased proportion of the rotational phase. An increase in rotational motion with cluster size has also been observed in simulations of flocks and experiments with fish schools~\cite{Tunstrom2013a}, suggesting that this mechanism for rotations may extend to systems other than just cell clusters. 

With increasing chemokine concentration gradient, the cell cluster spends an increased proportion of time in the running phase. An interesting consequence of this is that since the majority of rim/core exchanges take place while the cluster is in the running phase, we would expect the exchange to increase with increasing chemical gradient as well.  This is indeed what we see in the simulations (see SI-9) with a $50\%$ increase in exchange over a four-fold increase in gradient. These results are supportive of the conjectured functional benefit of exchanges of rim and core cells in maintaining robust chemotaxis ~\cite{Malet-Engra2014}.  An increase in rim/core exchange allows for the cells on the rim in high concentration gradients to shuffle back into the center of the cluster in order to replenish their chemical receptors, which become saturated while they are on the exposed rim of the cluster. At the same time, this brings core cells with unsaturated receptors to the rim allowing for more chemokine sensitivity. Thus, the cluster can utilize its collective dynamics to ensure a more robust response to gradients compared to individual cells. The emergence of exchanges between the periphery and interior, especially with increasing directional input at the periphery, might be of importance to flocks and swarms where sharing the inherent advantages/disadvantages of being at the core/rim (like temperature extremes in penguin colonies or the threat of predation in fish schools) is beneficial to the group as a whole.

Taken together, our results show that the rotations induced by rim-core coupling hold across a range of system sizes, propulsion strengths, noise values, and even in the presence of directional forcing.  They may even extend into three-dimensional rotations~\cite{Rorth2012,Miller2002,Bilder2012,Cai2016}, suggesting that the coupling between two swarming systems which are in different ordered phases can lead to interesting behaviors not seen in either system alone.  Heterogeneous behavior within a single group is a robust mechanism that cells or other types of swarming organisms may use to enhance rotating phases or other phases that would be unlikely or impossible to achieve otherwise. 

\color{white}
\cite{Ferrante2013}
\color{black}

\section*{Acknowledgements}
AG and KC were partially supported by National Science Foundation NSF grant EF-1038697 and NSF grant DMS-1616926, a James S. McDonnell Foundation Award and in part by the NSF-CREST: Center for Cellular and Bio-molecular Machines at UC Merced (NSF-HRD-1547848).  NSG gratefully acknowledges funding from the ISF (Grant No. 580/12).  Work in GS lab was partially supported by the Associazione Italiana per la Ri cerca sul Cancro (AI RC 10168); the 
Worldwide Cancer Research (AICR- 14-0335); from the European Research Council  (Advanced-ERC-268836)

\bibliographystyle{unsrt}


%

\end{document}